# Power Factor Correction of Inductive Loads using PLC

**Sayed Abdullah Sadat**
Member of Regime, National Load Control Center (NLCC)
Afghanistan's National Power Utility (DABS)
Sayed_abdullah@ieee.org

**E. Sreesobha**
Assistant Professor, Dept. of Electrical Engineering
UCE(A), Osmania University, Hyderabad, India.

**P.V.N. Prasad**
Professor, Dept of Electrical Engineering
UCE(A), Osmania University, Hyderabad, India.

**Abstract**

This paper proposes an automatic power factor correction for variable inductive loads, most dominantly induction motors (IM) utilizing the Programmable Logic Controllers (PLC). This hardware implementation of a 3Ø Inductive load system focuses on the automatic correction of power factor using PLC. With the help of PLC, different performance parameters – current level, real power and inductive power are obtained and logged in the PC. Using PLC program, according to control strategy to obtain a pre –specified power factor a set of capacitors sized in a binary rate will be switched on or off with the help of switching relays and contactors. This PLC control strategy relies on a lookup table which is prepared based on two input parameters - peak current and power factor, at constant voltage. From these parameters, PLC will calculate reactive power of the system and accordingly the right sequence of the capacitors are switched on in order to compensate reactive power.

## 1. Introduction

A Programmable logic controller (PLC) is a computer which is tailored specifically for a certain control tasks. With the invention of PLC, complex relay control systems are out dated. The advantage of PLC is without any system intervention, change of control strategy is possible. PLC is smaller, cheaper and more reliable than corresponding relay control systems. Improving power factor of an induction motor (IM) using PLC is presented in this paper. The very cause of poor power factor is the nature of the load. The inductive natured loads such as

Induction motors, arc lamps, electric discharge lamps and industrial heating furnaces cause low power factor. This low power factor will result in increased current magnitude, and additional losses of active power in the system. For a given constant power and voltage, the load current is always inversely proportional to the power factor. Lower the power factor, higher is the load current and vice-versa. Due to low power factor kVA rating of any equipment has to be increased. Since kVA rating of equipment is inversely proportional to power factor which will lead to increase in size of the equipment and cost. To facilitate increased current, conductor size has to be increased. Increased current causes more copper losses, poor efficiency and poor voltage regulation.

This paper proposes a laboratory model for a PLC based power factor correction to improve the power factor of inductive load system. Proposed algorithm calculates reactive power of induction motor load, compares it with existing capacitance reactive power in the capacitor bank and then compensating the system, so that system will not operate at leading power factor. Zero voltage switching of static switches will prevent the occurrence of the transients and harmonics.

No load, load and blocked rotor tests are conducted on the given 3Ø induction motor in the laboratory. Based on the experimental values plots of stator current, efficiency, torque and speed as a function of power factor are obtained. Active power, reactive power and power factors of motor are calculated at each load up to rated load. The variation of power factor is in the range of 0.24 to 0.41. From this information the size of the capacitors to be connected, so that power factor is improved, it is calculated by PLC automatically. In order to calculate this automatically, it is required to interface PLC with hardware through appropriate voltage and current magnitudes. Hence voltage and current signals are stepped down using current transformer (CT) and voltage transformer (VT).

## 2. Hardware Architecture

Hardware comprises of 3Ø supply, interfacing circuit, PLC, switching circuit, capacitor bank and 3Ø Induction Motor load as shown in fig.1.

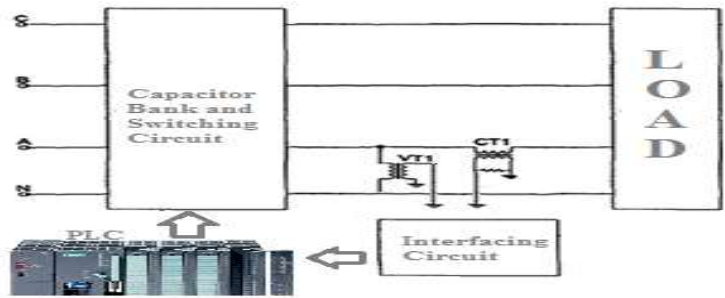

Fig.1 Block diagram of experimental set
### 2.1 Interfacing Circuit

The interfacing circuit consists of two main parts, namely Phase angle measuring circuit and current peak detector.

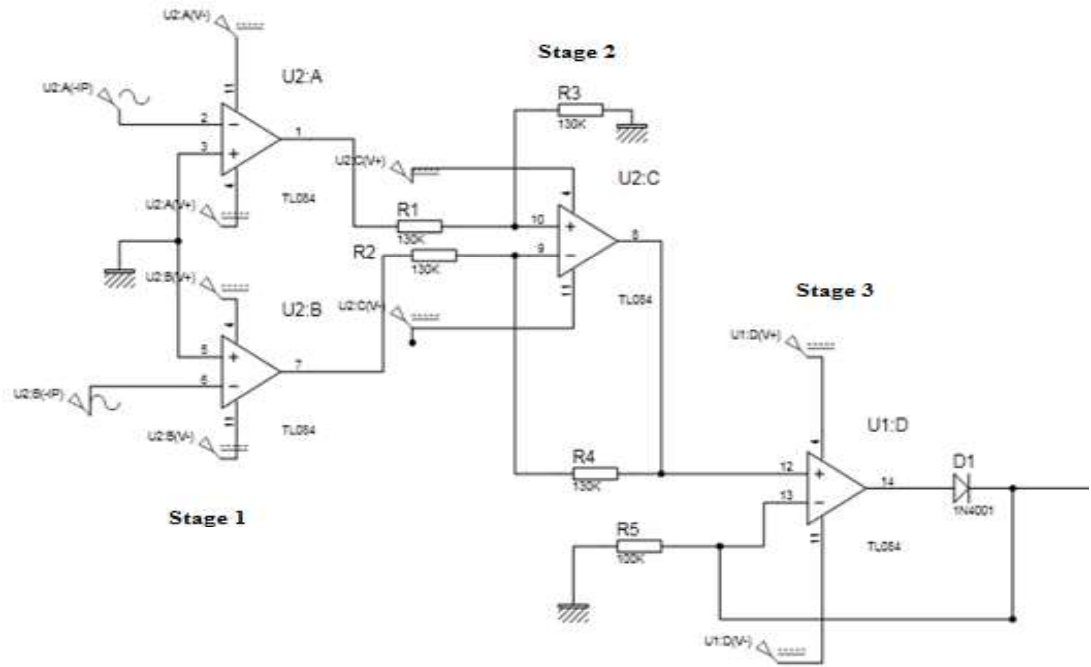

Fig.2 Circuit diagram for phase angle measurement

The phase angle measuring circuit which is shown in fig.2, comprises of converting stage, comparison stage, XOR stage and finally the clipping stage. At first stage the input sinusoidal current is converted into a square waveform using a comparator op-amp. Similarly the measured voltage is also converted to square waveform. Next the two square waveforms are then compared and the XOR logic is performed. At final stage by taking the output of the stage2, this stage will clip out all negative peaks present in the waveform. The output from port 6 as shown in fig.3 is given to the non-inverting terminal of the peak detector circuit. The conditioned output is a square wave. The amplitude of this square wave is trapped in the capacitor results in a signal, and this signal is given to PLC analog input port. The detected signal is scaled accordingly so as to obtain the desired results.

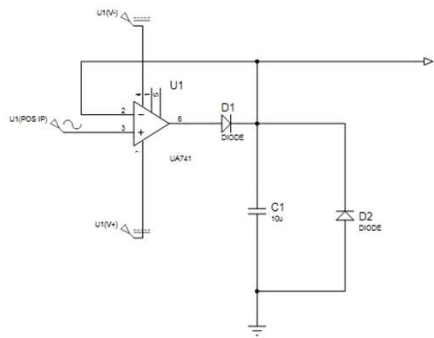

Fig.3 Circuit diagram of current peak detector

**2.2 Programmable Logic Controllers**

The power factor correcting circuit is driven by S7-300 PLC which is shown in fig.4, it consists of several modules – power supply, CPU, digital inputs, Digital outputs, and Analog –to-Digital converter. The digital input module is a 24 V DC, 13-30 V for "1" logic and -3 to 5 V for "0"logic, 32 input ports. The analog module is a 2-channel, 12-bit analog –to-digital converter (ACD). The digital output module is a 32 port, 0.5 A output current, 24 V DC rated load voltage [1].

The two outputs of the interfacing circuit are given to the PLC in the following way; the output of the phase angle measuring circuit is given to the digital input module of the PLC whereas the output of the current peak detector is given to the input of analog to digital converting module. The PLC then calculates the lagging reactive power of the system, and accordingly gives signal to digital output module. The digital output module has the switching circuit connected to it, which in turn connects the sequence of capacitors from capacitor bank which is shown in fig.1.

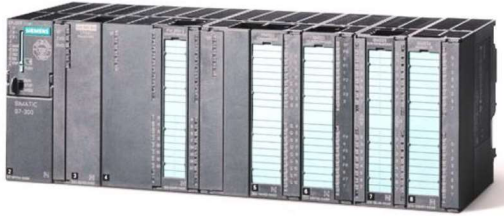

Fig.4 S7-300 PLC Siemens

### 2.3 Switching Circuits

The relay switches are energized directly by signal given out by digital output module of the PLC. By checking the bits of these two different sets of ports, it is possible to detect any switch failure. For more harmonic-free operation instead of relays the triac switches can be installed as shown in the fig.5. For this approach another VT is needed. This VT and its associated circuits provide zero crossing detection. The technique is useful to prevent the occurrence of transients, pseudo oscillation and harmonics. Moreover, three CTs and VTs, if employed independent phase compensations can also be achieved. This independent phase sensing of the lagging reactive power makes the controller suitable for an unbalanced power system.

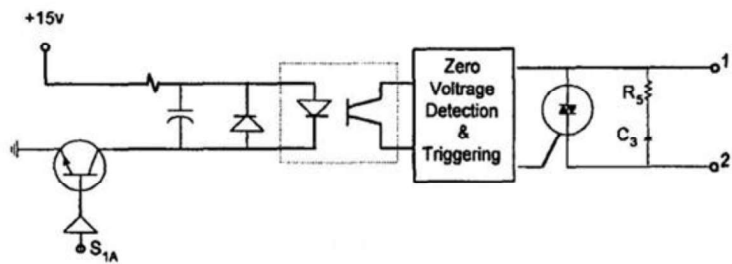

Fig 5: Circuit with Triac Switch Instead of relay

### 3. Testing

The prototype which is designed in the laboratory comprises of
a) Three Phase Induction Motor (IM): 3.7 kW, 1430 rpm, 50 Hz, 7.5 A, 415 V line voltage, a delta-connected 3-phase squirrel cage coupled with a brake test mechanism. b) S7-300 PLC as described in Section-2.2. c) Signal conditioning circuit to supply the measurement of lagging reactive power and the switching mechanism. d) Solid-state relays and three capacitors per phase for power factor compensation.

No load, load test and blocked rotor tests are conducted on the induction motor. From the load test results, power factor vs. load characteristics are plotted as shown in the Fig.6. [2]

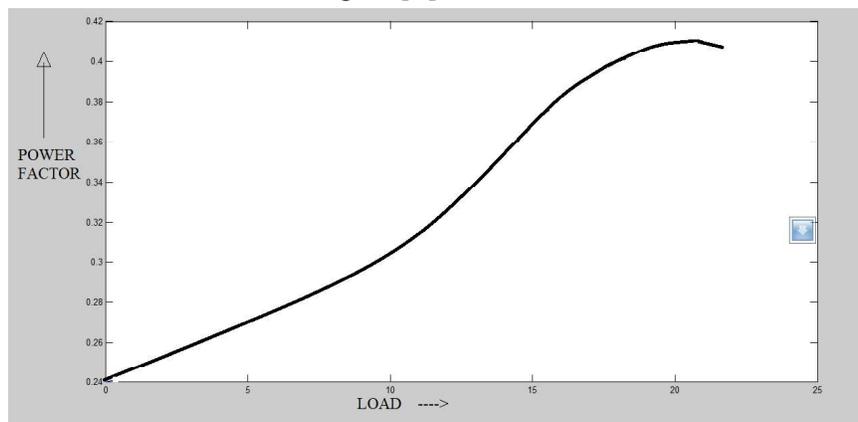

Fig.6 Power factor vs. load curve for given induction motor

From the characteristics, proper combination of capacitors which has to be connected into the circuit, for required VAr compensation is calculated. i.e. if the load current is below 3.9 A, 1 X 20 µF connected in delta and 2 X 2.5 µF capacitors has to be connected in to the circuit. If the load current is in between 3.9 A to 5.2 A then 2 X 2.5 µF connected in delta and 2X 20 µF connected in star has to be connected in to the circuit. The combination of three 20 µF star connected capacitors switched in for loads above 6 A to rated current. These three load current ranges are marked as "A", "B" and "C" in the fig.7. All these combinations are achieved by setting the appropriate relays on with PLC.

The variation of reactive power and power factor, obtained when loaded IM is controlled through PLC can be observed from the Table1 and Table 2. Improvement in the power factor of IM, with control strategy implemented by PLC can be observed from the values tabulated below.

Table 1: Variation of Power Factor without Compensation

| Supply(V) | Motor Current(A) | Speed(rpm) | P.F | Q(VAr) |
|---|---|---|---|---|
| 400 | 3 | 1447 | 0.24 | 2017.8 |
| 400 | 4 | 1467 | 0.28 | 2660.4 |
| 400 | 5 | 1465 | 0.37 | 3218.3 |
| 400 | 6 | 1446 | 0.40 | 3809.8 |
| 400 | 7 | 1441 | 0.41 | 4423.4 |

Table 2: Variation of Power Factor with Compensation

| Supply(V) | Motor Current(A) | Speed(rpm) | P.F | Q(VAr) |
|---|---|---|---|---|
| 400 | 3 | 1494 | 0.945 | 747.78 |
| 400 | 4 | 1474 | 1 | 0 |
| 400 | 5 | 1465 | 1 | 0 |
| 400 | 6 | 1453 | 0.99 | 586.41 |

## 4. Results

Square waves are obtained when sinusoidal voltage and current input are given to the two comparator op-amps. Comparison of these square waves shown in Fig.9 and Fig.11 will give the EXOR output. The resultant waveforms showing phase angle difference between voltage and current for different loads are shown in the Fig.8 and Fig.10 respectively. Improved phase angle relation and P.F can be examined easily from these voltage and current wave forms.

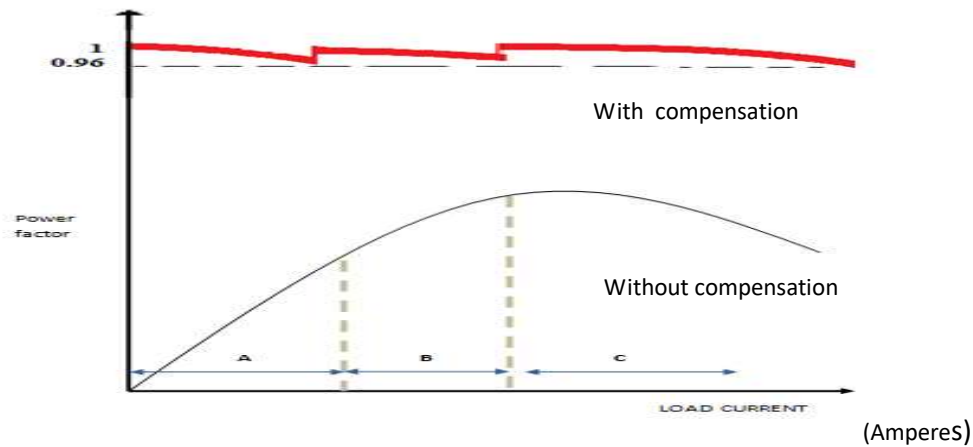

Fig.7 Power factor vs. load current

No load - without compensation:

**V=400 V, I=3 A, N=1447 rpm, P=500 W, pf=0.24lag**

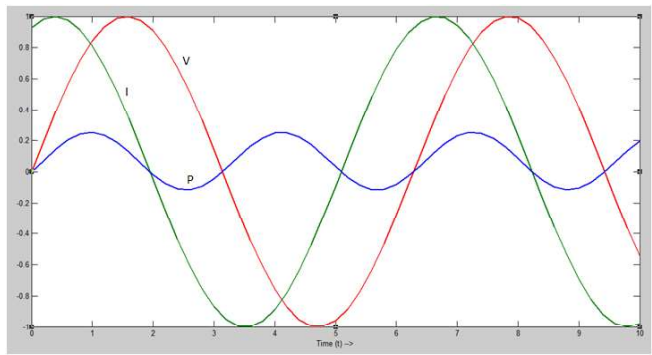

Fig.8 Voltage and Current wave forms for no load condition

On load - without compensation:

Load- 21kg brake load

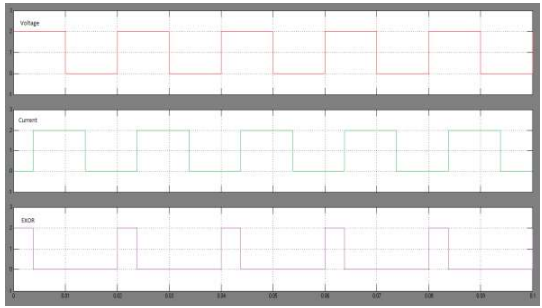 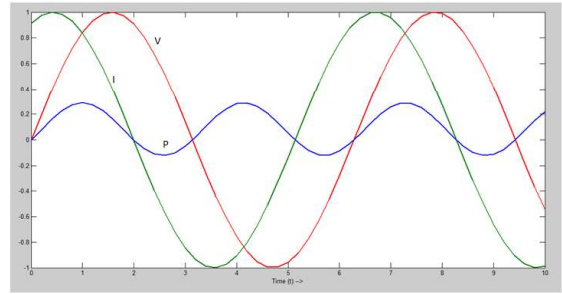

Fig.9 Comparator op-amps and EXOR Output wave forms:

Fig.10 Voltage and current wave forms

**V=400 V, I=7 A, N=1441 rpm, W=2000 w, pf=0.41**

On load - with compensation:

Load: 19.2Kg brake load

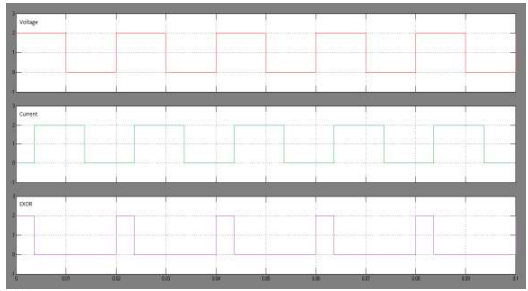 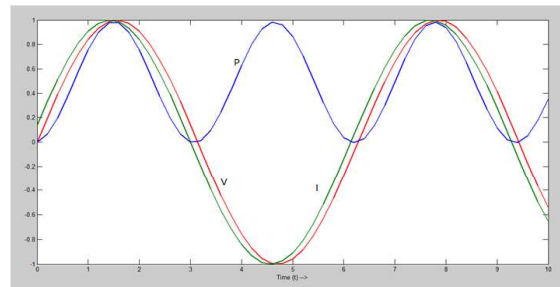

Fig.11 Comparator op-amps and EXOR Output wave forms:

Fig.12 Voltage and current wave forms – 19.2Kg brake load

## 5. Conclusions

The software strategy adopted will be particularly helpful in visualizing various problems involved in power factor correction. The salient features involved in the experiment are, (a) The utility of PLC in the design of an automatic power factor correction system, (b) The control strategy associated with automatic power factor correction system, (c) The physical visualization of the power factor improvement. Advantage of the proposed PFC compared with other controllers are that no drastic change in the hardware circuitry is envisaged to cope with different VAr ratings. Only the rating of the static switches and the capacitors needed are to be changed. Also there will be no change in the software strategy.